\documentclass[12pt]{article}  

\usepackage{epsf}  

\catcode `@=11
\@addtoreset{equation}{section} 
\catcode `@=12

\newcommand{\f}{\frac}
\newcommand{\half}{\frac12}

\newcommand{\beq}{\begin{eqnarray}}
\newcommand{\eeq}{\end{eqnarray}}

\def\sector#1#2{\ {\scriptstyle #1}\hskip 1mm  
\mathop{\square}\limits_{\lower 1mm\hbox{$\scriptstyle#2$}}\hskip 1mm}  
  
\def\tsector#1#2{\ {\scriptstyle #1}\hskip 1mm  
\mathop{\opensquare}\limits_{\lower 1mm\hbox{$\scriptstyle#2$}}^\sim\hskip 1mm}  

\def\){\right)}
\def\({\left( }
\def\]{\right] }
\def\[{\left[ }

\def\IC{{\relax\hbox{$\inbar\kern-.3em{\rm C}$}}}  
\def\IZ{{\relax\hbox{$\inbar\kern-.3em{\rm Z}$}}}

\def\al{\alpha}
\newcommand{\ap}{\alpha}

\newcommand{\ddd}{\cdot\cdot\cdot}

\def\be{\begin{equation}}
\def\ee{\end{equation}}
\def\bea{\begin{eqnarray}}
\def\eea{\end{eqnarray}}

\newcommand{\rem}[1]{}

\def\half{{1\over 2}}

\def\ltap{\ \raise.3ex\hbox{$<$\kern-.75em\lower1ex\hbox{$\sim$}}\ }
\def\gtap{\ \raise.3ex\hbox{$>$\kern-.75em\lower1ex\hbox{$\sim$}}\ }

\newcommand{\ba}{\begin{eqnarray}}
\newcommand{\ea}{\end{eqnarray}}
\newcommand{\no}{\nonumber \\}

\def\bea{\begin{eqnarray}}
\def\eea{\end{eqnarray}}

\def\half{\frac{1}{2}}

\begin{document}

\title{Condensation of Localized Tachyons \\ and Spacetime Supersymmetry}
\author{Soonkeon Nam$^\dagger$ and Sang-Jin Sin$^*$ \\ \\
\small \sl $^\dagger$Department of Physics and Research Institute 
of Basic Sciences, \\ \small \sl Kyung Hee University, Seoul, 
130-701, Korea\\ \small {\tt nam@khu.ac.kr}\\ \small \sl 
$^*$Department of Physics, Hanyang University, Seoul, 133-791, 
Korea\\ \small {\tt sjs@hepth.hanyang.ac.kr} 
 \\
} \maketitle
\begin{abstract}
We consider condensation of localized closed string 
tachyons by examining the recent proposal of Harvey, Kutasov, 
Martinec, and Moore. 
We first observe that the $g_{\rm cl}$ defined by HKMM does not reflect the space-time 
supersymmetry when the model has the SUSY. Especially for 
${\bf C}^2/{\bf Z}_N$ models, $g_{\rm cl}$ defined by them 
is highly peaked along the supersymmetric points in the space of orbifolds, which is unsatisfactory
property of the c-function  of the RG-flow.   
We give the modified definition of the $g_{\rm cl}$ in type II cases such that 
it has a valley along the  supersymmetric points in the orbifold moduli 
space. New definition predicts that the processes suggested by Adams,
Polchinski and Silverstein and was argued to be forbidden by  HKMM are in fact allowed. 
\end{abstract}

\newpage 
\section{Introduction}
Recently there has been a great interest on tachyon condensation in closed
string theories. Closed string tachyons indicate the decay of the spacetime
itself and full dynamical understanding is still lacking. 
To simplify the problem and to utilize the experience from the open string
case Adams, Polchinski, and Silverstein (APS) \cite{aps}
studied the cases where the tachyons are localized on a submanifold of 
the space-time, i.e, the tip of the orbifolds. 
They argued that  by the tachyon  condensation the  orbifolds decay and
the tip of the orbifold smooths out. They supported this conjecture by 
 D-brane probes  in sub-stringy regime and by general relativity analysis beyond string scale.
Since at the tip of the orbifold, the dimensionality of the space time is
less than full 10 dimensions, the rotation leaves no spinor invariant and
so supersymmetry is completely broken. Localized 
tachyons live in the twisted sector of the orbifold theory.
Vafa\cite{Vafa:2001ra} has argued for a similar picture in the context of gauged linear sigma
models and their mirror counterparts. Dabholkar and Vafa \cite{Dabholkar:2001wn}
proposed a closed string tachyon action where potential term was given by a c-function. 

In relation to this decay of orbifold background into smoother one, 
a very interesting proposal was put forward by Harvey, Kutasov, Martinec,
and Moore (HKMM) \cite{hkmm}, where they identified a certain quantity which decreases along
the RG flow, so that one may regard this quantity as a 
``potential'' governing the RG flow. 
From the world sheet point of view, 
localized closed string tachyons correspond to relevant perturbations in
the twisted sector of the world, and the decay induced by the tachyons
correspond to world sheet RG flow induced by the perturbation with tachyon operator.
They proposed to use the analog of the boundary entropy $g$ of Affleck and
Ludwig \cite{Affleck:tk}, and called this quantity $g_{\rm cl}$.
It can be obtained from the partition
function in the limit $\tau_{2}\to 0$, where $\tau_2$ is the imaginary
part of the modular parameter $\tau$:
\beq 
Z(\tau_2 \to 0) \sim  g_{\rm cl} \exp(\pi c/ 6\tau_2).
\eeq 
In conformal field theories, the central charge decreases along
RG flow 
\cite{Zamolodchikov:gt},\cite{Kutasov:1990sv}. 
However when the central charges are not
affected by RG flow, as with boundary
perturbations\cite{Affleck:tk}\cite{KMM}, 
the boundary entropy decreases along RG flow. 
The motive of HKMM was to introduce a similar
quantity in the case of localized tachyons. 
In this paper, we would like to  continue on this study of closed string
tachyon condensation by examining the proposal of HKMM. 
Our work is motivated from the observation that the $g_{\rm cl}$ is not 
reflecting the space-time supersymmetry when the model has the SUSY. 
In fact, for ${\bf C}^2/{\bf Z}_N$ models,  $g_{\rm cl}$ defined by HKMM 
is highly peaked along the supersymmetric points in the space of
orbifolds. 
Certainly this is not the property of the c-function of the RG-flow.
 
In this paper, we modify the definition of the $g_{\rm cl}$ in type II cases such that 
it has a valley along the  supersymmetric points in the orbifold moduli space.
The main technical differences of this paper with \cite{hkmm} are three folds:
\begin{enumerate}
  \item Instead of looking at the high temperature, we look at the low 
temperature.  We use 
 \be 
Z(\tau_2 \to \infty) \sim  g_{\rm cl} \exp(\pi c \tau_2/6),
\ee
to extract out the informations on the process due to  localized tachyon condensations. 
High and low temperature gives the same information for total degrees of freedom. 
But for localized (or delocalized) degrees of freedom separately, we get different results \cite{sin}. 
  \item Instead of looking at the twisted partition function 
  measuring localized degree of freedom, we look at the untwisted 
  partition function measuring the delocalized degrees of freedom.
   This leads us to the $g_{cl}$ which agrees with \cite{hkmm} for 
   type 0 case, and new $g^{II}_{cl}$ for type II case, which 
   guarantees the stability of the supersymmetric models. 
\item For type II case, we use full partition function rather than 
its bosonic part. For counting the central charges, it should not 
make much difference since the bosonic part does not contain the 
bulk tachyon anyway. But for the purpose of the getting $g_{cl}$, 
it is important to use the full partition function.

\end{enumerate}

At first looking,   considering the delocalized degrees of 
freedom to describe the localized tachyon condensation  may look odd.
However, in orbifold case, once the local geometry near the fixed 
point(s) is determined, the global geometry is also determined.
Once the local geometry changes, the global geometry must also 
change. So the it is natural  that the dynamics of delocalized degrees of freeedom 
should encodes the information of localized degree of freedom \cite{sin}.   
 
For technical convenience and for future interests, we consider superstring theory on Melvin 
background\cite{Me}. There are a few motivations for doing this: (i) it is exactly solvable
\cite{RT},   (ii) it has tachyons in the spectrum, 
and (iii) most importantly for our purpose, it reduces  to orbifolds \cite{TU} in a certain limit. 
Aspects of Melvin background in M-theory was studied in \cite{Costa:2000nw},
\cite{Gutperle:2001mb}, and \cite{Russo:2001tf}.
Tachyon condensation in this background was discussed using D-brane probe
\cite{Michishita:2001ph}, and in the context of
gauged linear sigma model and mirror symmetry \cite{David:2001vm}. 
For earlier literature on tachyon condensation in string theory see \cite{kogut}.

\section{ $g_{\rm cl}$, the potential for RG-flow}

The relevant orbifold partition function was first evaluated by Dabholkar in \cite{dabholkar}. 
Here for our technical convenience, we start from the 
expression of orbifold partition function given by Takayanagi and Uesugi in \cite{TU}
that is  obtained by taking a limit of Melvin background partition function, first calculated by Tyestlin and Russo \cite{RT}. 
We will follow closely the notations of Ref.\cite{RT} and Ref.\cite{TU}.

\subsection{${\bf C}/{\bf Z}_N$ Model }

Let us consider the path integral formulation of Green-Schwarz string on 
NS-NS Melvin background, $M_3\times R^{1,6}$.
$M_3$ is given by $S^1$ fibration over $R^2$. The coordinates of $R^2$ and
$S^1$  are
$(\rho, \phi)$ and $y$ with radius $R$. 
In the light-cone Green-Schwarz formulation, we have
eight bosonic fields $\rho, \phi, Y,X_i (i=2,\cdots,6)$.
We have 
\cite{RT}
\ba
Z(R,q,\beta)&=&(2\pi)^{-7}V_{7}R(\al')^{-5}\int \f{(d\tau)^2}{(\tau_2)^6}
\int (dC)^2 \sum_{w,w'\in {\bf Z}}\ 
\f{|\theta_{1}(\f{\chi}{2}|\tau)|^8}{|\eta(\tau)|^{18}
|\theta_{1}(\chi|\tau)|^2}\no
& &\times \exp\left[-\f{\pi}{\al\tau_2}(
4C\bar{C}-2\bar{C}R(w'-w\tau)+2CR(w'-w\bar{\tau}))\right], \label{PF1}
\ea 
where 
\ba
\chi=2\beta C+qR(w'-\tau w),\ \ \bar{\chi}=2\beta \bar{C}+qR(w'-\bar{\tau} w).
\ea
In the above $w, w'$ are the winding numbers, $q,\beta$ are the parameters
proportional to the strength of two gauge fields of the Kaluza-Klein Melvin
background. $C, \bar{C}$ are auxiliary parameters and $\tau = \tau_1+i\tau_2$ is the modular parameter.
Using   
\be
\theta_{3}(0|\tau)^3\theta_{3}(\chi|\tau)
-\theta_{2}(0|\tau)^3\theta_{2}(\chi|\tau)-
\theta_{4}(0|\tau)^3\theta_{4}(\chi|\tau)
=2\theta_{1}\left(\f{\chi}{2}|\tau\right)^4
\label{JA1},
\ee
and the  quasi-periodicity of theta-functions,  
the partition function (\ref{PF1}) 
in the  limit  $R\to 0$ and $\beta\al'/R \to 0$ with the rational value 
$qR={k \over N}$ finite, is given as follows \cite{TU};
\ba
&&\lim_{R\to 0}Z(R,q,\beta)
=(2\pi)^{-7}V_{7}R(\al)^{-4}\int \f{(d\tau)^2}{4(\tau_2)^5}
\sum_{l,m=0}^{N-1}\sum_{\ap,\beta\in {\bf Z}}\left(\lim_{R\to 0}
e^{-\f{\pi N^2R^2}{\al\tau_2}|\ap-\beta\tau|^2}\right)\no
&&\times\f{|\theta_3(\nu_{lm}|\tau)\theta_{3}(\tau)^3
-(-1)^{k\ap}\theta_2(\nu_{lm}|\tau)\theta_{2}(\tau)^3
-(-1)^{k\beta}\theta_4(\nu_{lm}|\tau)\theta_{4}(\tau)^3|^2}
{4|\eta(\tau)|^{18}
|\theta_{1}(\nu_{lm}|\tau)|^2}, \label{PF2}
\ea
where  $\nu_{lm}=\f{lk}{N}-\f{mk}{N}\tau$, and integers
$l,~m,~\alpha,~\beta$ are given as $w'=N\ap+l,w=N\beta+m$ 
$(l,m=0,1,\ddd,N-1)$. The case $l=m=0$ should be excluded, which correspond to the bosonic zero mode.

If $k$ is an odd integer, 
\be
Z(0,q,\beta)=\frac{1}{2}V_1V_{7}\int \f{(d\tau)^2}{4(\tau_2)}(4\pi^2\al\tau_2)^{-4}
\sum_{l,m=0}^{N-1} Z^{0}_{l,m}, \label{z0}
\ee
where 
\be
 Z^{0}_{l,m}=
 \f{|\theta_3(\nu_{lm}|\tau)\theta_{3}(\tau)^3|^2
+|\theta_2(\nu_{lm}|\tau)\theta_{2}(\tau)^3|^2
+|\theta_4(\nu_{lm}|\tau)\theta_{4}(\tau)^3|^2}
{2N|\eta(\tau)|^{18}
|\theta_{1}(\nu_{lm}|\tau)|^2},
\ee
and
$V_1=\lim_{R\to 0}\f{2\pi\al'}{NR}$ corresponds to the volume
of the noncompact direction. Thus the model was identified with the orbifold ${\bf C}/{\bf Z}_N$ in
type 0 string theory with radius $\f{\al'}{2NR}\to \infty$ \cite{TU}.
One should notice that the appearance of Type 0 GSO projection from Green Schwarz path-integral 
is a very surprising phenomena. It happens only for the orbifold limit and for the odd integer $k$, 
which is a measure zero set of moduli space of Melvin geometry \cite{tba}.  
 The sums over $l$ in 
(\ref{z0}) corresponds to ${\bf Z}_N$ projection (twists in time direction) and   
$m$ is over twisted sectors (twists in space direction), which is required for the modular invariance. 

Following Harvey et.al, we define the
untwisted partition function by 
\be
Z_{un}(\tau)= \sum_{l=0}^{N-1} Z_{l,0}
\ee
and twisted partition function by $Z_{tw}(\tau)=  
Z(\tau)-Z_{un}(\tau)$.
Now, we look at the low temperature limit of untwisted partition 
function. 
\be
Z^0_{un}(\tau \to \infty) = g^0_{cl} \cdot {\rm 
e}^{{ {2\pi}{\tau_2}}}, 
\ee
where 
\be  g_{\rm cl}^{0} =  \frac{1}{N}\sum_{l=1}^{N-1}\frac{1}{(2\sin[\frac{\pi k 
l}{N}])^{2}} ,
\ee
which reproduces the result first given by Harvey et.al 
\cite{hkmm} with somewhat opposite logic. 
Apart from the reproducing the result our procedure is more 
satisfying in the sense that the central charge is  counting the bulk degree of 
freedom, rather than the localized degrees of freedom.
Since we already explained in the introduction why bulk degrees of freedom contains the 
information of the process which is caused by the localized 
tachyon condensation, we do not repeat it here.

The sum can be explicitly evaluated to give
 \be
 g_{\rm cl}^{0} =\frac{1}{12}\left(N-\frac{1}{N}\right), 
 \ee
whose monotonicity in $N$ was used to support the conjecture that 
$g_{\rm cl}^{0} $ is the potential for the RG-flow \cite{hkmm}.

If $k$ is an even integer, 
\ba
Z(0,q,\beta)&=&V_1V_{7}\int \f{(d\tau)^2}{4\tau_2}(4\pi^2\al'\tau_2)^{-4} \sum_{l,m=0}^{N-1} Z^{II}_{l,m},\label{z1}
\ea
where 
\be
 Z^{II}_{l,m}=\f{|\theta_3(\nu_{lm}|\tau)\theta_{3}(\tau)^3
-\theta_2(\nu_{lm}|\tau)\theta_{2}(\tau)^3
-\theta_4(\nu_{lm}|\tau)\theta_{4}(\tau)^3|^2}
{4N|\eta(\tau)|^{18}
|\theta_{1}(\nu_{lm}|\tau)|^2}.
\ee
The model was identified with the orbifold ${\bf C}/{\bf Z}_N$ 
 in type II string theory \cite{TU}. 
These orbifolds include those discussed in \cite{aps} by APS ($k=N+1$).

For the definition of $g_{\rm cl}$ for type II, HKMM proposed that one should 
truncate the spectrum to bosonic sector,  namely,  one should make 
following change;
$$
(1+(-1)^{F_{L}})(1+(-1)^{F_{R}})/4 \to (1+(-1)^{F_{L}+F_{R}})/4.$$
The net result is that $g_{\rm cl}$ of HKMM is identical to that of type 0
up to a factor of half: 
\be
 g_{\rm cl, HKMM}^{II} = \frac{1}{2} g_{\rm cl}^{0} .
\ee
Although it has the right properties for many purposes, it has one 
critical  disadvantage; 
it does not reflect the 
consequence of the space-time supersymmetry of supersymmetric models. 
It means that $g_{\rm cl, HKMM}^{II}$ predicts that 
some of supersymmetric models can still decay.
 
Therefore, we proceed differently. We simply define $g_{\rm cl}^{II}$ 
as the leading term of the partition function of delocalized states,   
$Z_{un}(\tau_2\to\infty)$ as before. Due to the GSO projection the leading 
tachyon contribution cancels out.  
Notice that although the main tachyon contribution ($\sim \exp{(c\pi\tau_2/6)}$) 
cancels out, it is still non-zero. So we define 
this leading term as $g_{\rm cl}^{II},$ the potential for the 
RG-flow, namely
\be 
Z^{II}_{un}(\tau_2\to\infty) \sim g_{\rm cl}^{II}q^0.
\ee
A simple calculation shows that  
\be  g_{\rm cl}^{II} = \frac{1}{4N}\sum_{l=1}^{N-1} \frac{(2\sin[\frac{\pi k l}{2N}]  )^{8}}
{(2\sin[\frac{\pi k l}{N}])^{2}}.
\ee

 Notice that  $ Z^{0}_{un}(\tau_2\to \infty)$ is infinitely larger than $ Z^{II}_{un}(\tau_2\to\infty)$.
Since two objects do not share the same infinite factor,  it would be 
better to define $ Z_{un}(\tau_2\to \infty)$ itself as the potential of RG flow.
Under this criteria, there is no RG flow from type II to type 0,  \footnote{ 
Especially, the process of of decay of type II orbifold model into 
another type II orbifold plus baby universe of type 0, 
suggested in \cite{hkmm}, seems implausible.}
Within the type II orbifold models,  $g_{\rm cl}^{II}$ and 
$ g_{\rm HKMM, cl}^{II}$ gives the same physics, because their
 behaviours are very much the same: Both are monotonic in $N$ and 
the leading terms are linear in $N$.
The monotonicity of $ g_{\rm cl}^{II}$ can be easily checked using
following identity: 
\ba
 g_{\rm cl}^{II} &=& \frac{32}{3}\left(N+\frac{2}{N}\right) -30 \;\; {\rm for}\;\; 
 N>1 \no 
&=&0 \;\; {\rm for}\;\; N=1.
\ea
However, for higher dimensional orbifold models of ${\bf C}^2/{\bf Z}_N$, we 
will see that  $ g_{\rm cl}^{II}$ and 
$ g_{\rm HKMM cl}^{II}$  predict very different physics within type II theory. 

\subsection{${\bf C}^2/{\bf Z}_N$ Model}
We now consider the higher internal dimensional models.  We also start from 
the expression of orbifold partition function in Ref.\cite{TU}.
Here we have a background geometry of $M_5\times R^{1,4}$ where $M_5$ is a
fibration of $S^1$ over $R^2\times R^2$.
We now have two gauge fields $A_\phi$ and $A_\theta$ whose strengths involve
$q_1,q_2$ and $\beta_1,\beta_2$.
Again we will be considering the orbifold limit of $R\to 0$ with 
$\beta_i \al' /R \to 0$. We also have $q_i R = k_i/N$ $(i=1,2)$.

We first consider the case where $k_1+k_2$ is odd. 
The partition function is given by 
\ba
&&Z(0,q_1,q_2,\beta_1,\beta_2)=\half V_1V_7
\int \f{(d\tau)^2}{4\tau_2}
(4\pi^2\al'\tau_2)^{-3}
\sum_{l,m=0}^{N-1}\no
&&\times \f{|\theta_3(\nu^1_{l,m}|\tau)
\theta_3(\nu^2_{l,m}|\tau)\theta_3(\tau)^2|^2
+|\theta_2(\nu^1_{l,m}|\tau)\theta_2(\nu^2_{l,m}|\tau)\theta_2(\tau)^2|^2 
+|\theta_4(\nu^1_{l,m}|\tau)\theta_4(\nu^2_{l,m}|\tau)\theta_4(\tau)^2|^2}
{2N|\eta(\tau)|^{12}
|\theta_{1}(\nu^1_{l,m}|\tau)\theta_{1}
(\nu^2_{l,m}|\tau)|^{2}}.\no \label{PF23}
\ea
We have used $\nu^i_{l,m} = \frac{k_i}{N}(l-m\tau)$.
This corresponds to the type 0 case.
It is straightforward to see that 
\be  g_{\rm cl}^{0}(N,k_{1},k_{2})=
\frac{1}{N}\sum_{l=1}^{N-1}\frac{1}{(4\sin[\frac{\pi k_{1}l}{N}]
\sin[\frac{\pi k_{2}l}{N}])^{2}}.
\ee

Next we turn to the case when  $k_1+k_2$ is even:
\ba
&&Z(0,q_1,q_2,\beta_1,\beta_2)=V_1V_5
\int \f{(d\tau)^2}{4\tau_2}
(4\pi^2\al'\tau_2)^{-3}
\sum_{l,m=0}^{N-1}\no
&&\times \f{|\theta_3(\nu^1_{l,m}|\tau)
\theta_3(\nu^2_{l,m}|\tau)\theta_3(\tau)^2
-\theta_2(\nu^1_{l,m}|\tau)\theta_2(\nu^2_{l,m}|\tau)\theta_2(\tau)^2 
-\theta_4(\nu^1_{l,m}|\tau)\theta_4(\nu^2_{l,m}|\tau)\theta_4(\tau)^2|^2}
{4N\ |\eta(\tau)|^{12}
|\theta_{1}(\nu^1_{l,m}|\tau)\theta_{1}
(\nu^2_{l,m}|\tau)|^{2}}.\no \label{PF22}
\ea
This is the type II case and we have the following expression:
\be  g_{\rm cl}^{II}(N,k_{1},k_{2})=\frac{1}{4N}\sum_{l=1}^{N-1} \frac{(4\sin[\frac{\pi
(k_{1}+k_{2})l}{2N}]\ \sin[\frac{\pi(k_{1}-k_{2})l}{2N}]\ )^{4}}
{(4\sin[\frac{\pi k_{1}l}{N}]\sin[\frac{\pi k_{1}l}{N}])^{2}} .
\label{aps}
\ee
Here again, the main tachyonic piece is cancelled out, and we 
defined $g_{\rm cl}^{II}$ as the leading piece of 
$Z_{un}(\tau_2\to \infty)$ as before.
Notice that along the $k_1=\pm k_2$, the $g_{\rm cl}^{II}$ vanishes and 
have a valley reflecting the supersymmetry of the model. We can clearly see
this in Fig. 1. 
This guarantees the stability of the supersymmetric models.
On the other hand, according to HKMM, we have  $ g_{\rm cl}^{II}=\half
g_{\rm cl}^{0}$, which 
shows a ridge instead of a valley along the supersymmetric 
points  $k_1=\pm k_2$, as we can  see in  Fig. 2. 

We give a contour plot of $ g_{\rm cl}(N=23,k_{1},k_2)$ in figures 1 and 2.
The lower the value of $g_{\rm cl}$ the darker the shade is in the figures.

\begin{figure}[htbp1]
 \epsfxsize=80mm
\centerline{\epsfbox{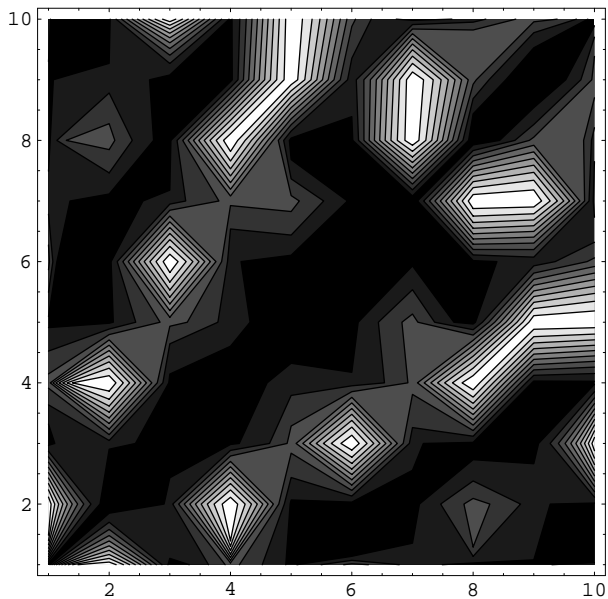}}
 \caption{Contour plot of $g_{\rm cl}^{II}$ in $(k_1,k_2)$ plane. Only
$k_1>0,k_2>0$ is shown. It has a valley along the line $k_1=k_2$.
Actually the axes correspond to $k_i$-th prime numbers.}
  \label{Fig1}
\end{figure}
 
\begin{figure}[htbp2]
 \epsfxsize=80mm
\centerline{\epsfbox{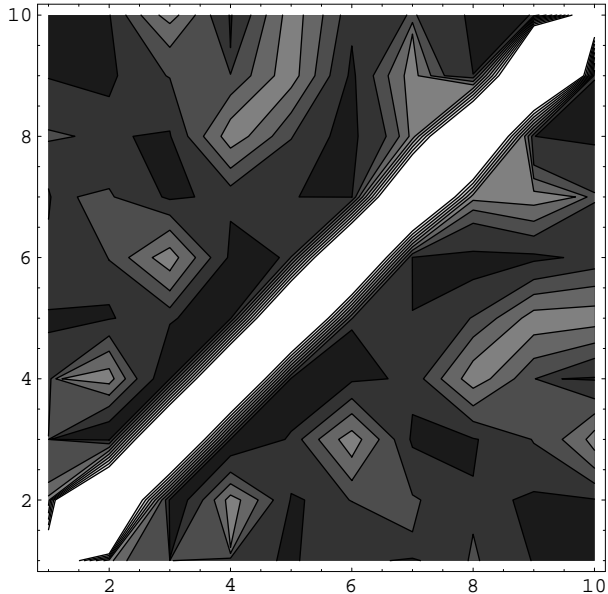}}
 \caption{Contour plot of $g^{II}_{\rm cl,HKMM}=\half g_{\rm cl}^{0}$. 
 It has a ridge along the diagonal $k_1=k_2$, implying the instability of supersymmetric models. }
  \label{Fig2}
\end{figure}

Another example that is worthwhile mentioning is the example of APS: 
${\bf C}^2/{\bf Z}_{2l(3)} \to {\bf C}^2/{\bf Z}_{l(1)}$. This was mentioned 
 as a possible counter example of $g_{\rm cl, HKMM}^{II}$ in Ref.\cite{hkmm}.
Based on the following inequality: 
\be 
 g_{\rm cl}^{0}(2l,3,1) <  g_{\rm cl}^{0}(l,1,1),
\ee
HKMM argued that the process should be impossible.
However, according to $ g_{\rm cl}^{II}$ given in Eq.(\ref{aps}), this 
process is possible, 
since we now have the following inequality:
\footnote{ In fact $g_{\rm cl}^{II}(2l,3,1) >  g_{\rm cl}^{II}(l,1,1)+ g_{\rm cl}^{II}(l,-3,1)$ also holds.}
\be
 g_{\rm cl}^{II}(2l,3,1) >  g_{\rm cl}^{II}(l,1,1).
\ee

\subsection{Large $N$ limit}
It may be of some interests to notice that there are well defined 
limit of large $N$ limits of $g_{\rm cl}$ that can be given by 
integral expressions.
For ${\bf C}/{\bf Z}_N$ model, there is no $k$ dependence in finite 
$N$, so one may expect that it would not be interesting to take $N \to \infty$ limit.
However it is not so clear whether this is true for the limiting 
cases. So we write down the expressions here.
For ${\bf C}/{\bf Z}_N$ models, we have
\be
g^0_{\rm cl}(k)=\frac{1}{\pi}\int^{\pi}_0\frac{dx}{(2\sin[ k 
x])^{2}} 
\ee
and
\be
g_{\rm cl}^{II}(k)= \frac{1}{4\pi}\int_{0}^{\pi} dx \frac{(2\sin[k x/2])^{8}}
{(2\sin[k x])^{2}} 
\label{aps1}
\ee
For ${\bf C}^2/{\bf Z}_N$ models we have, 
\be
g^0_{\rm cl}(k_1,k_2)=\frac{1}{\pi}\int^{\pi}_0\frac{dx}{(4\sin[ k_1x]\sin[ k_2 x])^{2}}.
\ee
One should notice that this is singular when $k_1=\pm k_2$. 
The integrand  has $1/x^4$ singularity near 
$x = n\pi/k <\pi $  and it corresponds to  $N^3$ dependence for finite $N$.   
For type II case we have,
\be  
g_{\rm cl}^{II}(k_{1},k_{2})= \frac{1}{4\pi}\int_{0}^{\pi} dx \frac{(4\sin[
(k_{1}+k_{2})x/2]\ \sin[(k_{1}-k_{2})x/2]\ )^{4}}
{(4\sin[ k_{1}x]\sin[ k_{2}x])^{2}}. 
\label{aps2}
\ee
Here also we can easily see the valley along the lines $k_1=\pm k_2$.
We believe that these are related to the generic (non-orbifold) case of 
Melvin background \cite{tba}.  

\section{Discussion}
 
 In this paper, we examined recent proposal made in Ref.\cite{hkmm}, namely, the interpretation of the high energy spectral density
 $g_{\rm cl}$ as a Tachyon potential. 
We have shown that the $g_{\rm cl}$ defined by HKMM does not reflect the space-time 
supersymmetry when the model has the SUSY. Especially for 
${\bf C}^2/{\bf Z}_N$ models, $g_{\rm cl}$ defined by them 
is shown to be highly peaked along the supersymmetric points in the space of orbifolds.
This clearly is unsatisfactory property of the ``potential'' of the RG-flow.   
We gave the modified definition of the $g_{\rm cl}$ in type II cases such that 
it has a valley along the  supersymmetric points $k_1=\pm k_2$.
Our new definition suggests that the  process suggested by APS and was argued 
to be forbidden by HKMM should  in fact be allowed. 

Now we list some of the issues to be studied in future. 
One may want to extend the analysis of D-brane probe and that of HKMM to the orbifold models corresponding to $k_2\ne 1$,
which involve the orbifolds that goes beyond Hirzebruch-Jung geometry. 
Another issue is the discussion of large $N$ behavior of 
$g_{\rm cl}$ calculated in this paper. 
It seems to be related to the generic Melvin 
background.  We wish to come back to these issues in future publication. 

{\bf Acknowledgements}  \\
We would like to thank David Kutasov for pointing out an error in the first version of this paper.
The work of SJS is supported by the research fund of Hanyang University (HY-2000).  
The work of SN is supported by Korea Research Foundation Grant KRF-2001-041-D00049.
This work is also supported by BK21 program of Korea Research Fund (2001).

\end{document}